\documentclass[a4paper,11pt]{article}
\usepackage{pos}

\hypersetup{pdfauthor={Reinhold Kaiser},
            pdftitle={Approaching the Continuum Limit of the Deconfinement Critical Point for \texorpdfstring{$\nf=2$}{Nf=2} Staggered Fermions},
            pdfsubject={QCD Phase Diagram},
            pdfkeywords={LQCD, Deconfinement Phase Transition, Columbia Plot},
            pdfcreator={Latex with hyperref},
}

\usepackage{units}
\usepackage{tikz}
\usetikzlibrary{
	external,
	shapes.geometric,
	arrows, backgrounds,
	positioning,
	calc,
	patterns,
}
\usepackage{pgfplots}
\usepackage{pgfplotstable}
\usepgfplotslibrary{fillbetween}
\usepgfplotslibrary{groupplots}
\pgfplotsset{
compat=1.16,
cmhplot/.style={color=black,mark=none,line width=1pt,<->},
soldot/.style={color=black,only marks,mark=*},
every tick label/.append style={font=\scriptsize},
legend style={font=\scriptsize},
/pgf/declare function={
		l(\m,\ns,\a,\mc)=1.604+\a*(1/\m-1/\mc)*\ns^(1/0.6301);
	},
	/pgf/declare function={
		lc(\m,\ns,\a,\mc,\b)=(1.604+\a*(1/\m-1/\mc)*\ns^(1/0.6301))*(1+\b*\ns^(-0.8940));
	},
	select coords between index/.style 2 args={
    	x filter/.code={
        	\ifnum\coordindex<#1\fi
        	\ifnum\coordindex>#2\fi
    	}
	},
}

\usepackage{wrapfig}
\usepackage{physics}
\usepackage{bbold}
\usepackage{csquotes}
\usepackage{bm}
\usepackage{subcaption}

\newcommand{\clqcd}{\texttt{CL\kern-.25em\textsuperscript{2}QCD}}
\newcommand{\nf}{N_\text{f}}
\newcommand{\bpc}{\beta_{\text{pc}}}
\newcommand{\mc}{m_{\text{c}}}
\newcommand{\rmass}{\tilde m}
\newcommand{\rb}{\tilde\beta}

\title{Progress on the QCD deconfinement critical point for $\nf=2$ staggered fermions}
\ShortTitle{The deconfinement critical point for $N_\text{f}=2$ staggered fermions}

\author*[a,b]{Reinhold Kaiser}
\author[a,b]{Owe Philipsen}

\affiliation[a]{Institute for Theoretical Physics - Goethe University,\\
  Max-von-Laue-Str. 1, 60438 Frankfurt am Main, Germany}

\affiliation[b]{John von Neumann Institute for Computing (NIC), GSI,\\
Planckstr. 1, 64291 Darmstadt, Germany}

\emailAdd{kaiser@itp.uni-frankfurt.de}
\emailAdd{philipsen@itp.uni-frankfurt.de}

\abstract{The global center symmetry of quenched QCD at zero baryonic chemical potential is broken spontaneously at a critical temperature $T_c$ leading to a first-order phase transition. Including heavy dynamical quarks breaks the center symmetry explicitly and weakens the first-order phase transition for decreasing quark masses until it turns into a smooth crossover at a $Z(2)$-critical point. We investigate the $Z(2)$-critical quark mass value towards the continuum limit for $\nf=2$ flavors using lattice QCD in the staggered formulation. As part of a continued study, we present results from Monte-Carlo simulations on $N_\tau=8,10$ lattices. Several aspect ratios and quark mass values were simulated in order to obtain the critical mass from a fit of the Polyakov loop to a kurtosis finite size scaling formula. Moreover, the possibility to develop a Ginzburg-Landau effective theory around the $Z(2)$-critical point is explored.}

\FullConference{%
The 39th International Symposium on Lattice Field Theory,\\
8th-13th August, 2022,\\
Rheinische Friedrich-Wilhelms-Universität Bonn, Bonn, Germany
}

\begin{document}
\maketitle
\section{Introduction}
The thermal transition of Quantum Chromodynamics (QCD) at zero baryon chemical potential and physical quark masses has been determined as an analytic crossover~\cite{aoki06}.
The QCD phase diagram as a function of the baryon chemical potential and the temperature $T$ can not be explored directly by lattice QCD for non-zero real chemical potentials due to a severe sign problem.
However, exploring the phases of QCD for unphysical parameter values, in particular unphysical quark masses, with first-principle methods gives valuable insights into the rich phase structure of QCD.

\begin{wrapfigure}[18]{r}{0.5\textwidth}
\includegraphics[width=0.5\textwidth]{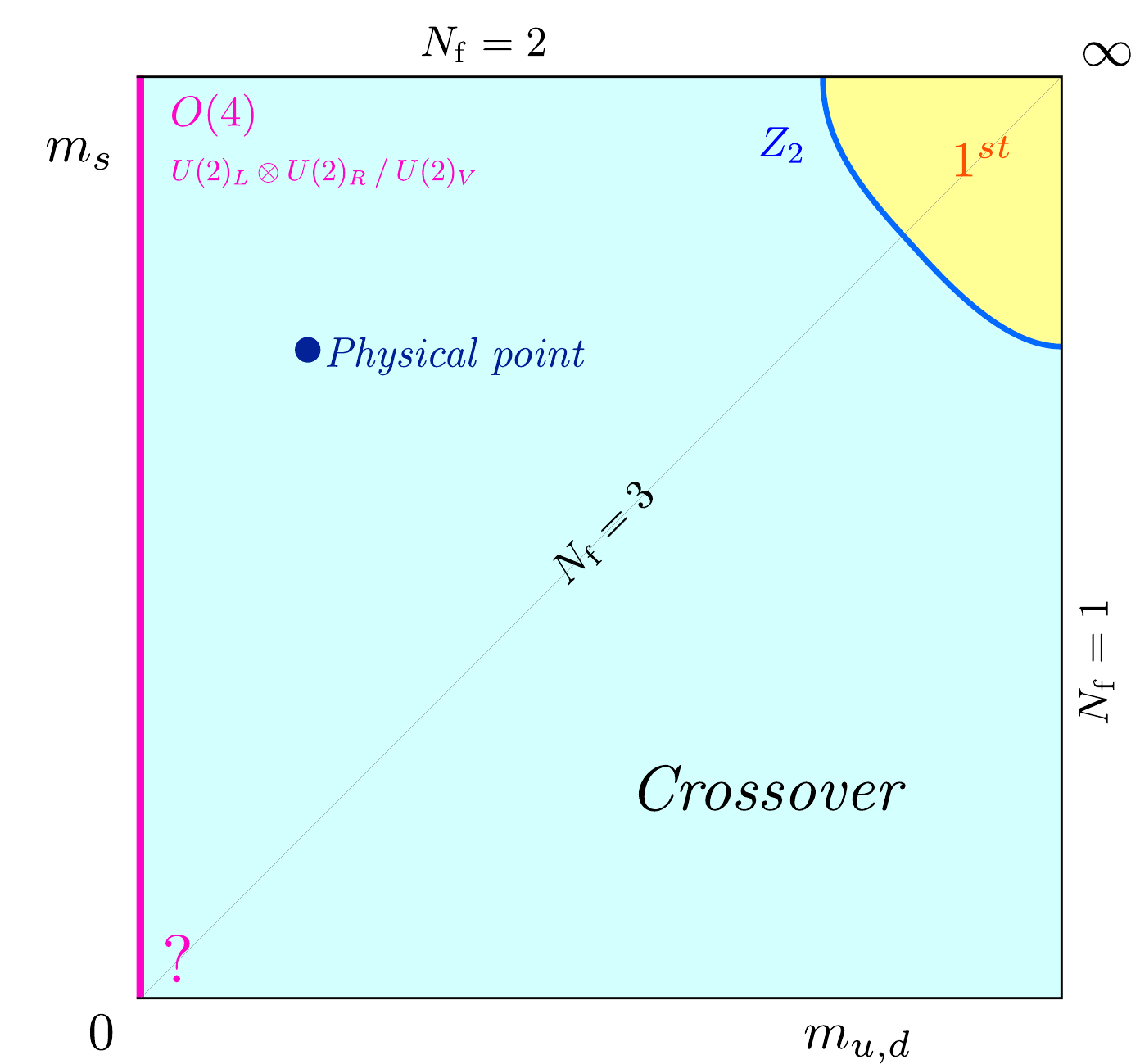}
\caption{The Columbia plot for the second-order scenario of the chiral phase transition, from ref.~\cite{cuteri21}. \label{fig:columbia-plot}}
\end{wrapfigure}

The Columbia plot (see fig.~\ref{fig:columbia-plot}) determines the nature of the QCD thermal transition as a function of the degenerate up- and down-quark mass $m_{u,d}$ and the strange quark mass $m_s$ at zero chemical potential~\cite{brown90}.
In contrast to the light quark mass region, where strong evidence exists, that there is no chiral transition of first order in the continuum limit~\cite{cuteri21}, the first order region in the heavy corner is known to persist from investigations of pure gauge theory~\cite{boyd96}.
The deconfinement transition in the heavy quark regime, which we focus on in this work, is related to the spontaneous breaking of the $Z(3)$ center symmetry.
For infinitely heavy quark masses the center symmetry of QCD is exact.
When including heavy dynamical quarks, the center symmetry is broken explicitly and the first-order phase transition weakens with decreasing quark masses.
At the $Z(2)$ second-order boundary the first order transition turns into a smooth, analytic crossover.

The location of the $Z(2)$-critical point for $\nf=2$ quark flavors has been investigated using the Wilson fermion action for three different lattice spacings $a$~\cite{cuteri20}.
Here we present results for the same $Z(2)$-critical point employing the staggered fermion action on lattices with two different lattice spacings.
As a continuation of the work from ref.~\cite{kaiser22} final results from the coarser lattice are shown, as well as new preliminary results from the finer lattice.
Another point on the $Z(2)$-critical boundary has been studied also with staggered fermions for $\nf=3$ quark flavors~\cite{borsanyi21}.

The resulting critical quark masses will provide a first-principles benchmark for effective theories, that are not excluded from investigating non-zero real chemical potentials.
Examples are effective lattice theories of QCD employing the hopping parameter expansion~\cite{saito11,saito14,ejiri19,ejiri20,kiyohara21,wakabayashi22,kanaya22,fromm12,fromm13,aarts16} and Polyakov loop models in the continuum~\cite{fischer15,lo14}.
Additionally, a first attempt is made to determine an effective Ginzburg-Landau theory around the deconfinement critical point as a function of the staggered lattice QCD parameters.
Similar as in ref.~\cite{hatta03}, the pseudo-critical phase boundary can be determined as a function of the lattice QCD parameters, which allows to fit it to data originating from lattice QCD simulations.

\section{Simulation Details}
The Monte Carlo importance sampling simulations are conducted on Euclidean lattices with dimensions $N_\sigma^3\times N_\tau$ employing the standard Wilson gauge action and the unimproved staggered fermion action (for detailed formulations see ref.~\cite{gattringer10}).
The lattice QCD parameters are the inverse gauge coupling $\beta(a)=6/g^2$ and the bare quark mass $am$.
The inverse gauge coupling $\beta$ controls the lattice spacing $a$ and tunes the temperature through the relation $T=(a(\beta)N_\tau)^{-1}$.
Increasing $N_\tau$ while keeping $T$ constant leads to the continuum limit, where $N_\tau=8,10$ for our simulations.
To account for the thermodynamic limit, simulations at several aspect ratios $LT=N_\sigma/N_\tau\in\{4,5,6,7,8\}$ are performed, where $L$ is the spatial extent of the thermodynamic system.
In order to localize the transition point, $2$ to $4$ $\beta$ values are simulated around the pseudo-critical $\bpc$.
For the quark mass $am$, 5 or 6 values around the critical quark mass $a\mc$ at the $Z(2)$-critical point are simulated.
For each set of parameters, 4 independent Markov chains are produced, which are thermalized before with a sufficient number of updating steps.

The gauge configurations are generated according to the RHMC algorithm, which is implemented in the Open-CL based lattice QCD code \clqcd~\cite{sciarra21a}.
It is run on the GPU-clusters L-CSC at the GSI in Darmstadt and on the Goethe-HLR at the Center for Scientific Computing in Frankfurt.
The huge number of simulations is efficiently handled and monitored using the bash tool \texttt{BaHaMAS}~\cite{sciarra21b}.

\section{Analysis of the Deconfinement Transition}
The order parameter associated with the deconfinement phase transition is the volume averaged Polyakov loop 
\begin{equation}
L =\frac{1}{N_\sigma^3}\sum_{\bm n}\frac{1}{3}\Tr\left[\prod_{n_4=0}^{N_\tau-1}U_4(\bm n, n_4)\right],
\end{equation}
with temporal gauge links $U_4(n)$.
The standardized moments of the absolute value of the Polyakov loop
\begin{equation}
B_n = \frac{\expval{\left(\abs{L}-\expval{\abs{L}}\right)^n}}{\expval{\left(\abs{L}-\expval{\abs{L}}\right)^2}^{n/2}}.
\end{equation}
contain information about the shape of its distribution.
We analyze the skewness $B_3$ as a function of $\beta$ in order to determine the transition point
at $\bpc$ through the symmetry condition $B_3(\bpc)=0$.
The kurtosis $B_4$ at $\bpc$ determines the order of the transition, where $B_4$ assumes universal values in the thermodynamic limit:
$B_4(\bpc)=1$ for a first-order phase transition, $B_4(\bpc)=3$ for a crossover and $B_4(\beta_\text{c})=1.604(1)$~\cite{blote95} for a $Z(2)$ second-order phase transition in three dimensions.
Both the skewness and the kurtosis are interpolated as a function of $\beta$ using the multiple histogram reweighting method~\cite{ferrenberg89}.
Hence, a precise estimate of $\bpc$ allows to obtain a good estimate of $B_4(\bpc)$.
To account for the volume dependency of $B_4(\bpc; N_\sigma, am)$, a finite size scaling analysis gives
\begin{equation}\label{equ:kurtosis-finite-size}
B_4(\bpc;N_\sigma, am)=\left(1.604+B x + 
\order{x^2}\right)\cdot \left( 1 + C N_\sigma^{y_t-y_h} + \order{N_\sigma^{2(y_t{-}y_h)}}\right)
\end{equation}
depending on the scaling variable $x=\left(\frac{1}{am}-\frac{1}{a\mc}\right)N_\sigma^{1/\nu}$~\cite{jin17}.
$y_t=1/\nu=1.5870(10)$ and $y_h=2.4818(3)$~\cite{pelissetto02} are Ising 3D universal exponents.
$B$ and $C$ are constants that remain to be determined and the correction term $C N_\sigma^{y_t-y_h}$ can be neglected for sufficiently large volumes.
The critical mass $a\mc$ for a certain $N_\tau$ can then be extracted as a fit parameter from a fit of formula~\eqref{equ:kurtosis-finite-size} to the $B_4(\bpc;N_\sigma,am)$ data.

\section{Simulation Results}
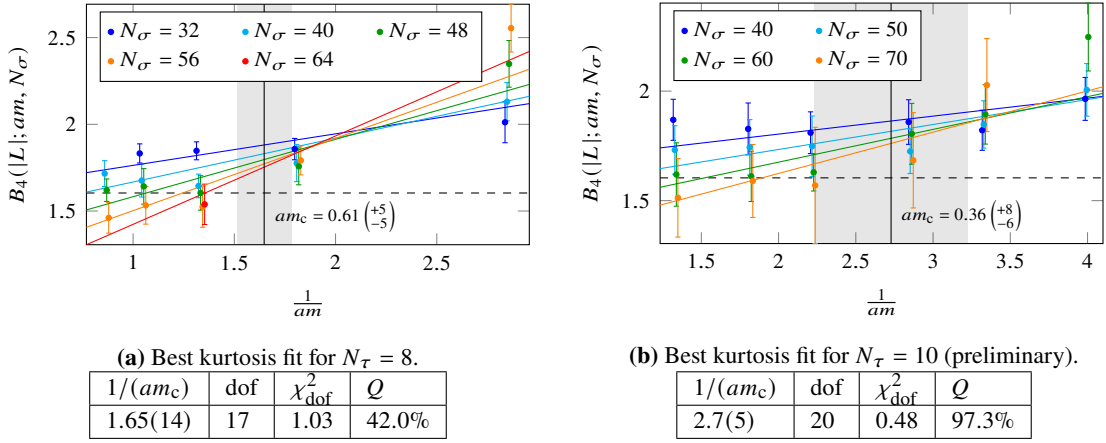
\begin{figure}
\captionsetup[subfigure]{justification=centering}
\centering
\begin{subfigure}[b]{0.49\textwidth}
\centering
\begin{tikzpicture}[every mark/.append style={mark size=1pt}, every error bar/.append style={mark size=1pt}]
\pgfplotstableread{anc/kurtosis-points-nt8.dat}{\kurtosistable};
\begin{axis}[width=\textwidth, height=4.77cm, enlargelimits=false, legend pos=north west, legend columns=3, legend style={/tikz/every even column/.append style={column sep=0.5cm}}, xlabel=\scriptsize $\frac{1}{am}$, ylabel={\scriptsize $B_4(\abs{L}; am, N_\sigma)$}, clip=false]
\path[fill=black,draw=none, opacity=0.1]
(axis cs:1.783,\pgfkeysvalueof{/pgfplots/ymin}) --
(axis cs:1.783,\pgfkeysvalueof{/pgfplots/ymax}) -- 
(axis cs:1.512,\pgfkeysvalueof{/pgfplots/ymax}) --
(axis cs:1.512,\pgfkeysvalueof{/pgfplots/ymin}) -- cycle;
\addplot [black, dashed, domain=1/1.15-0.1:1/0.35+0.1, forget plot]{1.604}; 
\addplot [
	blue,
	select coords between index={0}{3},
	only marks,
	mark size=0.5mm,
	error bars/.cd, y dir = both, y explicit,
] table [x expr=1/\thisrowno{0}-0.02, y=Kurtosis, y error=errorKurtosis] {\kurtosistable};
\addplot [
	cyan,
	select coords between index={4}{8},
	only marks,
	mark size=0.5mm,
	error bars/.cd, y dir = both, y explicit,
] table [x expr=1/\thisrowno{0}-0.01, y=Kurtosis, y error=errorKurtosis] {\kurtosistable};
\addplot [
	green!60!black,
	select coords between index={9}{13},
	only marks,
	mark size=0.5mm,
	error bars/.cd, y dir = both, y explicit,
] table [x expr=1/\thisrowno{0}, y=Kurtosis, y error=errorKurtosis] {\kurtosistable};
\addplot [
	orange,
	select coords between index={14}{18},
	only marks,
	mark size=0.5mm,
	error bars/.cd, y dir = both, y explicit,
] table [x expr=1/\thisrowno{0}+0.01, y=Kurtosis, y error=errorKurtosis] {\kurtosistable};
\addplot [
	red,
	select coords between index={19}{19},
	only marks,
	mark size=0.5mm,
	error bars/.cd, y dir = both, y explicit,
] table [x expr=1/\thisrowno{0}+0.02, y=Kurtosis, y error=errorKurtosis] {\kurtosistable};
\legend{$N_\sigma=32$, $N_\sigma=40$, $N_\sigma=48$, $N_\sigma=56$, $N_\sigma=64$};
\addplot [blue, domain=1/1.15-0.1:1/0.35+0.1]{lc(1/x, 32, 0.0006341, 0.607, 3.827)};
\addplot [cyan, domain=1/1.15-0.1:1/0.35+0.1]{lc(1/x, 40, 0.0006341, 0.607, 3.827)};
\addplot [green!60!black, domain=1/1.15-0.1:1/0.35+0.1]{lc(1/x, 48, 0.0006341, 0.607, 3.827)};
\addplot [orange, domain=1/1.15-0.1:1/0.35+0.1]{lc(1/x, 56, 0.0006341, 0.607, 3.827)};
\addplot [red, domain=1/1.15-0.1:1/0.35+0.1]{lc(1/x, 64, 0.0006341, 0.607, 3.827)};

\draw (axis cs:1.6474283086845711,\pgfkeysvalueof{/pgfplots/ymin}) -- (axis cs:1.6474283086845711,\pgfkeysvalueof{/pgfplots/ymax});
\node[anchor = west] at (axis cs: 1.65, 1.47) {\tiny $a\mc=0.61\left(^{+5}_{-5}\right)$};

\end{axis}
\end{tikzpicture}
\caption{Best kurtosis fit for $N_\tau=8$.\\
\begin{tabular}{|l|l|l|l|}
\hline
$1/(a\mc)$ &  $\text{dof}$ & $\chi^2_{\text{dof}}$ & $Q$\\
\hline
$1.65(14)$ & $17$ & $1.03$ & $42.0\%$\\
\hline
\end{tabular}
}
\end{subfigure}
\hfill
\begin{subfigure}[b]{0.49\textwidth}
\begin{tikzpicture}[every mark/.append style={mark size=1pt}, every error bar/.append style={mark size=1pt}]
\pgfplotstableread{anc/kurtosis-points-nt10.dat}{\kurtosistable};
\begin{axis}[width=\textwidth, height=4.77cm, enlargelimits=false, legend pos=south east, legend columns=2, legend style={/tikz/every even column/.append style={column sep=0.5cm}}, xlabel=\scriptsize $\frac{1}{am}$, ylabel={\scriptsize $B_4(\abs{L}; am, N_\sigma)$}, clip=false, legend pos=north west]
\path[fill=black,draw=none, opacity=0.1]
(axis cs:3.225668841,\pgfkeysvalueof{/pgfplots/ymin}) --
(axis cs:3.225668841,\pgfkeysvalueof{/pgfplots/ymax}) -- 
(axis cs:2.230845454,\pgfkeysvalueof{/pgfplots/ymax}) --
(axis cs:2.230845454,\pgfkeysvalueof{/pgfplots/ymin}) -- cycle;
\addplot [black, dashed, domain=1/0.75-0.1:1/0.25+0.1, forget plot]{1.604}; 
\addplot [
	blue,
	select coords between index={0}{5},
	only marks,
	mark size=0.5mm,
	error bars/.cd, y dir = both, y explicit,
] table [x expr=1/\thisrowno{0}-0.015, y=Kurtosis, y error=errorKurtosis] {\kurtosistable};
\addplot [
	cyan,
	select coords between index={6}{11},
	only marks,
	mark size=0.5mm,
	error bars/.cd, y dir = both, y explicit,
] table [x expr=1/\thisrowno{0}-0.005, y=Kurtosis, y error=errorKurtosis] {\kurtosistable};
\addplot [
	green!60!black,
	select coords between index={12}{17},
	only marks,
	mark size=0.5mm,
	error bars/.cd, y dir = both, y explicit,
] table [x expr=1/\thisrowno{0}+0.005, y=Kurtosis, y error=errorKurtosis] {\kurtosistable};
\addplot [
	orange,
	select coords between index={18}{22},
	only marks,
	mark size=0.5mm,
	error bars/.cd, y dir = both, y explicit,
] table [x expr=1/\thisrowno{0}+0.015, y=Kurtosis, y error=errorKurtosis] {\kurtosistable};
\legend{$N_\sigma=40$, $N_\sigma=50$, $N_\sigma=60$, $N_\sigma=70$};
\addplot [blue, domain=1/0.75-0.1:1/0.25+0.1]{lc(1/x, 40, 0.00020252895147949033, 1/2.7282571475309316, 4.37736892030209)};
\addplot [cyan, domain=1/0.75-0.1:1/0.25+0.1]{lc(1/x, 50, 0.00020252895147949033, 1/2.7282571475309316, 4.37736892030209)};
\addplot [green!60!black, domain=1/0.75-0.1:1/0.25+0.1]{lc(1/x, 60, 0.000202528951479490339, 1/2.7282571475309316, 4.37736892030209)};
\addplot [orange, domain=1/0.75-0.1:1/0.25+0.1]{lc(1/x, 70, 0.00020252895147949033, 1/2.7282571475309316, 4.37736892030209)};

\draw (axis cs:2.7282571475309316,\pgfkeysvalueof{/pgfplots/ymin}) -- (axis cs:2.7282571475309316,\pgfkeysvalueof{/pgfplots/ymax});
\node[anchor = west] at (axis cs: 2.7282571475309316, 1.43) {\tiny $a\mc=0.36\left(^{+8}_{-6}\right)$};

\end{axis}
\end{tikzpicture}
\caption{Best kurtosis fit for $N_\tau=10$ (preliminary).\\
\begin{tabular}{|l|l|l|l|}
\hline
$1/(a\mc)$ & $\text{dof}$ & $\chi^2_{\text{dof}}$ & $Q$\\
\hline
$2.7(5)$  & $20$ & $0.48$ & $97.3\%$\\
\hline
\end{tabular}
}
\centering
\end{subfigure}
\caption{Kurtosis data and fits for $N_\tau=8,10$.
The data points are shifted due to readability. The colored lines show the combined fit, indicating the corresponding volume by the color.
The dashed line indicates the infinite volume kurtosis value for the $Z(2)$ second-order transition.
The vertical black line and the grey band localize the critical mass and its error. 
 \label{fig:kurtosis-fits}}
\end{figure}

Fitting the kurtosis from formula~\eqref{equ:kurtosis-finite-size} including the relevant correction term gives critical quark mass values $a\mc$ for $N_\tau=8,10$.
The fits and the used data points can be seen in fig.~\ref{fig:kurtosis-fits} together with tables summing up the fit result.
The effect of the correction term on the fit is recognized as pairwise different crossing points of the kurtosis lines for different volumes, that are shifted upwards with respect to the infinite volume kurtosis crossing point (crossing of the black solid and black dashed line).
Larger error bars on the kurtosis points of $N_\tau=10$ in comparison with $N_\tau=8$ can be explained by larger autocorrelation times for the larger lattices and still low statistics for the larger $N_\sigma$ for $N_\tau=10$.
The bad fit quality, indicated by the large $Q$-parameter, and the large error on the critical mass are a consequence of the large error bars on the kurtosis points.
This issue will disappear with increasing statistics.

\begin{table}[h]
\centering
\begin{tabular}{|l|l|l|l|l|l|l|l|}
\hline
$N_\tau$ & $am$ & $\beta_{c}$ & $am_\pi$ & $a\; \{\unit{fm}\}$ & $m_\pi\; \{\unit{GeV}\}$ & $T_c\; \{\unit{MeV}\}$ & $m_\pi/T_c$ \\
\hline
$8$ & $0.6070$ & $5.9940$ & $1.79918(9)$ & $0.0885(9)$ & $4.01(4)$ & $278(3)$ & $14.39$\\
$10$ & $0.35$ & $6.0828$ & $1.38085(15)$ & $0.0691(7)$ & $3.95(4)$ & $286(3)$ & $13.81$\\
\hline
\end{tabular}
\caption{Pion mass measurement and scale setting results. The data for $N_\tau=10$ is preliminary for $am=0.35$, $a\mc$ is not yet final.\label{tab:results}}
\end{table}

\begin{figure}
\centering
\begin{tikzpicture}
    \begin{axis}[
      name=ax1,
      xticklabel style={/pgf/number format/fixed},
      xmin=0, xmax=0.13,
      ymin=3.85, ymax=5.1,
      x dir=reverse,
      width=5cm,
      height=5cm,
      axis x line=box,
      axis y line=left,
      x axis line style={-},
      extra x ticks={0.06915, 0.0880},
      extra x tick labels={10, 8},
      extra x tick style={tick pos=top, ticklabel pos=top},
      xtick pos=left,
      legend style={at={(axis cs:0.01, 5)}, anchor=north east},
      ylabel=\scriptsize $m_\pi^{Z(2)}\; \{GeV\}$,
    ]
      \addplot[
  red, mark options={red}, only marks, mark size=0.3ex,
  error bars/.cd, 
    y fixed,
    y dir=both, 
    y explicit
] table [x=x, y=y,y error=error, col sep=comma, row sep=crcr] {
    x,  y,       error\\
    0.0691, 3.95, 0.04\\
    0.0885, 4.01, 0.04\\
    };
      \addlegendentry{\scriptsize{staggered}};
      \addplot [black] coordinates{(0,0) (0,6)};
    \end{axis}
    
    \begin{axis}[
      xticklabel style={/pgf/number format/fixed},
      width=5cm,
      height=5cm,
      xmin=0, xmax=0.13,
      ymin=3.85, ymax=5.1,
      yticklabel pos=right,
      at={(ax1.south east)},
      anchor=south west,
      axis x line=box,
      axis y line= right,
      x axis line style={-},
      extra x ticks={0.0691, 0.0876, 0.1186},
      extra x tick labels={10, 8, 6},
      extra x tick style={tick pos=top, ticklabel pos=top},
      xtick pos=left,
      legend style={at={(axis cs:0.01, 5)}, anchor=north west},
    ]
      \addplot[
  blue, mark options={blue}, only marks, mark size=0.3ex,
  error bars/.cd, 
    y fixed,
    y dir=both, 
    y explicit
] table [x=x, y=y,y error=error, col sep=comma, row sep=crcr] {
    x,  y,       error\\
    0.0691, 4.39, 0.05\\
    0.0876, 4.51, 0.05\\
    0.1186, 5.01, 0.05\\
    };
    \addlegendentry{\scriptsize{Wilson}};
    \end{axis}
    \node[anchor=north, yshift=-1.5ex] at (ax1.south east) {\scriptsize $a\; \{\text{fm}\}$};
    \node[anchor=south, yshift=1.5ex] at (ax1.north east) {\scriptsize $N_\tau$};
\end{tikzpicture}
\caption{Comparison with the results from Wilson fermions~\cite{cuteri20}. The staggered value for $m_\pi^{Z(2)}$ at $N_\tau=10$ is preliminary for $am=0.35$.}\label{fig:pion-masses}
\end{figure}
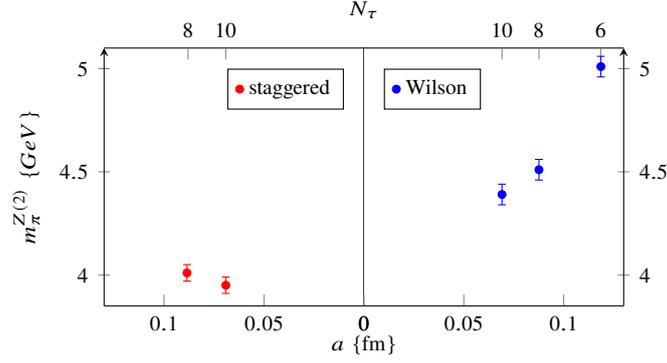

The scale is set via the $w_0$ scale~\cite{borsanyi12}, which is based on the Wilson flow~\cite{luescher10}.
The pion mass in physical units is compared to results from Wilson fermions~\cite{cuteri20} in fig.~\ref{fig:pion-masses}.
The error bars represent the error from the pion mass measurement.
The fit error on $a\mc$, which is significantly larger, is not shown in the plot.
The complete results of the pion mass measurement and the scale setting are presented in table~\ref{tab:results}.
For both $N_\tau$, the critical pion mass in lattice units $am_\pi^{Z(2)}$ is larger than 1, implying that the pion, represented by its Compton wavelength, is not yet resolved by the lattice.

\section{Development of an Effective Ginzburg-Landau Theory}
We investigate the possibility to find an effective Ginzburg-Landau theory in the vicinity of the deconfinement critical point that describes the lattice data.
To determine an ansatz for the Landau functional, the $Z(3)$-symmetric combinations of the order parameter, which is the complex Polyakov loop, have to be found.
Invariant combinations of $L$ and its complex conjugate $L^*$ include $L^*L=\abs{L}^2$, $(L^*)^3$, $L^3$ and $(L^*L)^2=\abs{L}^4$ up to fourth order, which appear in the general Landau functional as
\begin{equation}
\mathcal L(L, L^*) = B_2\abs{L}^2+B_{3,*}(L^*)^3 + B_3 L^3 +B_4 \abs{L}^4.
\end{equation}
In the thermodynamic limit $L$ is restricted to one center sector or the spontaneously broken phase, such that $\ev{L^3}=\ev{(L^*)^3}=\ev{\abs{L}^3}$.
Including dynamical fermions, the center symmetry is explicitly broken, where $1/(am)$ is the symmetry breaking field.
In the thermodynamic limit $\ev{\Re(L)}\rightarrow\ev{\abs{L}}$ for the volume averaged Polyakov loop.
With these assumptions we substitute $\abs{L}\equiv\Lambda$ to reduce the problem to a one-component field.

The general Landau functional is then defined as
\begin{equation}
\mathcal L(\Lambda; \rb,\rmass)=C_2(\rb,\rmass)\Lambda^2+C_3(\rb,\rmass)\Lambda^3+C_4(\rb,\rmass)\Lambda^4-\frac{1}{am}\Lambda,
\end{equation}
where the coefficients $C_i$ depend on the reduced quantities $\rb=\frac{\beta}{\beta_c}-1$ and $\rmass=\frac{\mc}{m}-1$, which become 0 at the critical point.
To leading order the coefficients are
\begin{equation}
C_i(\rb,\rmass)\approx c_{i,0}+c_{i,\beta}\rb+c_{i,m}\rmass,
\end{equation}
with constants $c_{i,0}$, $c_{i,\beta}$ and $c_{i,m}$.
 
At a second order phase transition with $(\beta_c, a\mc)$, the three conditions for the Landau functional at the critical point~\cite{hatta03} are
\begin{align}
\mathcal L'(\Lambda; 0, 0)&=0\label{equ:critical-condition1}\\
\mathcal L''(\Lambda; 0, 0)&=0\label{equ:critical-condition2}\\
\mathcal L'''(\Lambda; 0, 0)&=0,\label{equ:critical-condition3}
\end{align}
where eq.~\eqref{equ:critical-condition1} demands that the Landau free energy is minimized.
Eq.~\eqref{equ:critical-condition2} represents the fact that the curvature has to become zero at a second order phase transition.
By eq.~\eqref{equ:critical-condition3} we impose a minimum for the curvature $\mathcal L''(\Lambda; 0,0)$ of the Landau functional at the critical point. $L''''(\Lambda;0;0)>0$ always, since $c_{4,0}>0$ for a global minimum of $\mathcal L$ to exist.
The solution to the equation system is
\begin{align}
\Lambda_c &=\frac{3}{2\cdot c_{2,0}\cdot a\mc}\label{equ:lambda-c}\\
c_{3,0} &= -\frac{4\cdot c_{2,0}^2\cdot a\mc}{9}\\
c_{4,0} &= \frac{2\cdot c_{2,0}^3\cdot (a\mc)^2}{27}.
\end{align}

To reduce the number of unknown parameters, we assume $C_3$ and $C_4$ to be constant and set $C_2(\rb)=c_{2,0}+c_{2,\beta}\rb$.
This leads to a minimal ansatz for the Landau functional
\begin{equation}\label{equ:minimal-functional}
\mathcal{L}_{\text{min}}(\Lambda;\rb,\rmass)=(c_{2,0}+c_{2,\beta}\rb)\Lambda^2-\frac{4\cdot c_{2,0}^2\cdot a\mc}{9}\Lambda^3+\frac{2\cdot c_{2,0}^3\cdot (a\mc)^2}{27}\Lambda^4-\frac{1}{am}\Lambda,
\end{equation}
which is only valid for sufficiently small $\rmass$ and $\rb$.
This minimal ansatz allows to determine a simple functional form of the phase boundary $\bpc(am)$ in the vicinity of the critical point.
\begin{wrapfigure}[15]{r}{0.4\textwidth}
\centering
\begin{tikzpicture}[every mark/.append style={mark size=1pt}, every error bar/.append style={mark size=1pt}]

\pgfplotstableread{anc/betaPcs_nt8_ns56.dat}{\betaPcsTable};
\pgfplotstableread{anc/etaPcs_nt8_ns56.dat}{\etaPcsTable};

\def\ctwo{174.256}
\def\amc{0.607}
\def\betac{5.99343}
\def\ctwobeta{-3436.87}

\begin{axis}[width=0.4\textwidth, height=4cm, enlargelimits=true, ylabel={\scriptsize $\Lambda_{\text{pc}}(am)$}, clip=false, legend pos=north west, at={(0, 3cm)}]
\addplot [
	black,
	only marks,
	mark size=0.8mm,
	mark=+,
	error bars/.cd, y dir = both, y explicit,
] table [x expr=1/\thisrowno{0}, y=etaPc, y error=errorEtaPc] {\etaPcsTable};
\addplot[
	domain=1/1.15:1/0.35,
	blue,
] {3/2/\amc/\ctwo};
\end{axis}
\begin{axis}[width=0.4\textwidth, height=4cm, enlargelimits=true, xlabel=\scriptsize $\frac{1}{am}$, ylabel={\scriptsize $\bpc(am)$}, clip=false, legend pos=north west,]
\addplot [
	black,
	only marks,
	mark size=0.8mm,
	mark=+,
	error bars/.cd, y dir = both, y explicit,
] table [x expr=1/\thisrowno{0}, y=betaPc, y error=errorBetaPc] {\betaPcsTable};
\addplot[
	domain=1/1.15:1/0.35,
	blue,
	samples=200,
] {\betac+(\amc*x-1)*\ctwo*\betac/3/\ctwobeta};
\end{axis}
\end{tikzpicture}
\caption{Combined fit of eqs.~\eqref{equ:lambda-pc},~\eqref{equ:beta-pc}  to data from simulations on $N_\tau=8$, $N_\sigma=56$ lattices. \label{fig:landau-fit}}
\end{wrapfigure}
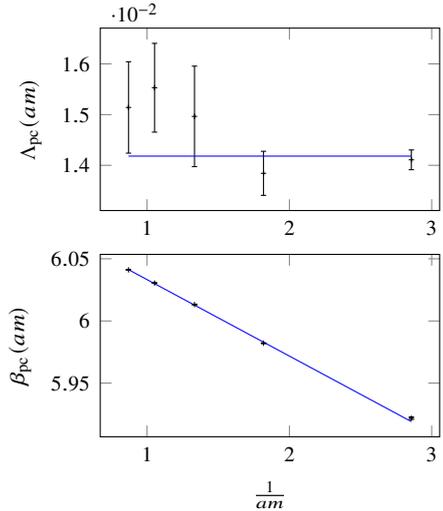
Solving the equation system
\begin{align}
\mathcal L'(\Lambda; \rb, \rmass)&=0\\
\mathcal L'''(\Lambda; \rb, \rmass)&=0,
\end{align}
for $\Lambda$ and $\rb$, the pseudo-critical quantities
\begin{align}
\Lambda_{\text{pc}} &= \frac{3}{2\cdot c_{2,0}\cdot a\mc}\label{equ:lambda-pc}\\
\rb_{\text{pc}} &= \frac{c_{2,0}}{3\cdot c_{2,\beta}}\rmass\label{equ:beta-pc}
\end{align}
are obtained.
The curvature $\mathcal L''(\Lambda; \rb, \rmass)$ must be minimal again, but it is not always $0$ anymore, which is valid for polynomials up to fourth order.

The unknown coefficients can be determined by fitting eqs.~\eqref{equ:lambda-pc}~and~\eqref{equ:beta-pc} to data from simulations and using the critical mass value from the kurtosis fit of one specific $N_\tau$.
The resulting combined fit for $N_\tau=8$, $N_\sigma=56$ is shown in fig.~\ref{fig:landau-fit}, with fit parameters $\beta_\text{c}=5.9934(4)$, $c_{2,0}=174(4)$ and $c_{2,\beta}=-3.44(9)\cdot 10^3$ that minimize the reduced $\chi^2_{\text{dof}}=3.69$.

$\Lambda$ is non-zero at the critical point, such that we can shift the Landau functional and express it in terms of a new order parameter $\eta=\Lambda-\Lambda_\text{c}$, which becomes $0$ at the critical point.
The shifted minimal Landau functional is
\begin{equation}
\mathcal L_\text{min}(\eta;\rb,\rmass)=-\left(\frac{1}{am}-\frac{1}{a\mc}\left(1+\frac{3c_{2,\beta}}{c_{2,0}}\rb\right)\right)\eta +c_{2,\beta}\rb\eta^2+\frac{2\cdot c_{2,0}^3\cdot (a\mc)^2}{27}\eta^4,
\end{equation}
where the symmetry breaking field now depends both on $am$ and on $\beta$.

\section{Conclusions}
This work presents progress on determining the deconfinement critical point for $\nf=2$ staggered fermions in the continuum limit.
Results from two different lattice spacings for $N_\tau=8,10$ have been shown.
The task to perform the continuum limit is challenging mainly due to increasing autocorrelation times.
The preliminary fit for $N_\tau=10$ is not yet well constrained, as the kurtosis errors are still large for currently running simulations.
Despite these challenges, it is necessary to simulate at even larger $N_\tau$, because the pion is still not resolved by $N_\tau=10$ lattices.
To circumvent the issue of large computational cost, an effective Ginzburg-Landau theory in the vicinity of the deconfinement critical point has been explored in a first step.
Based on a minimal Landau functional, the functional form of the phase boundary is fitted to data from simulations.
However, further investigations are required, which includes taking into account the $N_\tau$ dependence of the Landau functional.
This would allow to study the continuum limit with the effective Ginzburg-Landau theory.

\acknowledgments{
The authors acknowledge support by the Deutsche Forschungsgemeinschaft (DFG, German Research Foundation) through the CRC-TR 211 \enquote{Strong-interaction matter under extreme conditions} – project number 315477589 – TRR 211 and by the State of Hesse within the Research Cluster ELEMENTS (Project ID 500/10.006).
We thank the Helmholtz Graduate School for Hadron and Ion Research (HGS-HIRe) for its support as well as the staff of L-CSC at GSI Helmholtzzentrum für Schwerionenforschung and the staff of Goethe-HLR at the Center for Scientific Computing Frankfurt for computer time and support.
}

\bibliographystyle{JHEP}
\bibliography{main}

\end{document}